\documentclass[superscriptaddress,amsmath,amssymb,aps,pra,twocolumn,floatfix,longbibliography]{revtex4-1} %
\usepackage{bm}
\usepackage{graphicx}
\usepackage{float}
\usepackage{placeins}
\usepackage{braket}
\usepackage{bbold}
\usepackage[colorlinks, linkcolor=mycol, citecolor=mycol, urlcolor=mycol, breaklinks]{hyperref}
\usepackage{xcolor}

\usepackage{mathtext}

\definecolor{mycol}{RGB}{10,55,130}

\newcommand  {\supp} {\mathrm{sup}}
\newcommand {\at} {\mathrm{at}}

\newcommand {\inphyni} {Universit\'e C\^ote d'Azur, CNRS, Institut de Physique de Nice, 06560 Valbonne, France}

\begin{document}

\title{Superradiance as single scattering embedded in an effective medium}
\author{P. Weiss}
\altaffiliation[Present address: ]{Physikalisches Institut, Eberhard Karls Universit\"at T\"ubingen, T\"ubingen, Germany}
\affiliation{\inphyni}
\author{A. Cipris}
\affiliation{\inphyni}
\author{R. Kaiser}
\affiliation{\inphyni}
\author{I. M. Sokolov}
\affiliation{Department of Theoretical Physics, Peter the Great St.-Petersburg Polytechnic University, 195251, St.-Petersburg, Russia}
\affiliation{Institute for Analytical Instrumentation, Russian Academy of Sciences, 198103, St.-Petersburg, Russia}
\author{W. Guerin}
%\email{william.guerin@inphyni.cnrs.fr}
\affiliation{\inphyni}

\begin{abstract}
We present an optical picture of linear-optics superradiance, based on a single scattering event embedded in a dispersive effective medium composed by the other atoms. This linear-dispersion theory is valid at low density and in the single-scattering regime, i.e. when the exciting field is largely detuned. The comparison with the coupled-dipole model shows a perfect agreement for the superradiant decay rate. Then we use two advantages of this approach. First we make a direct comparison with experimental data, without any free parameter, and show a good quantitative agreement. Second, we address the problem of moving atoms, which can be efficiently simulated by adding the Doppler broadening to the theory. In particular, we discuss how to recover superradiance at high temperature.
\end{abstract}

\date{\today}

\maketitle

\section{Introduction}

Superradiance generally refers to the accelerated radiation rate of excited atoms due to the collective interaction of the sample with light and the vacuum reservoir. It has been originally introduced by Dicke for a collection of excited atoms in a small volume \cite{Dicke:1954}, and experimentally studied with large-size, low-density samples \cite{Feld:1980, Gross:1982}. Although Dicke's approach is based on collective atomic states, it is also interesting to develop an optical picture of the superradiance (or superfluorescence \cite{Bonifacio:1975}) phenomenon, in particular to understand propagation effects associated to the size and shape of the sample. In the case of many excited atoms, one may consider superradiance as the transient form of stimulated emission \cite{Feld:1980}.

More recently the subject of `single-photon superradiance' has been brought up by Scully \textit{et al.} \cite{Scully:2006, Scully:2009} and experimentally observed using weak continuous excitation of large-size and dilute samples \cite{Araujo:2016,Roof:2016}. In this linear-optics regime the picture of stimulated emission obviously cannot apply. Is it still possible, then, to use an optical picture of superradiance in that case?

In this article we present such an optical picture, which is based on a single scattering event embedded in the effective medium built by the whole atomic sample. The effective medium has a complex refractive index that introduces attenuation and dispersion. In this picture, the physics of linear-optics superradiance appears to be very close to the one of optical precursors and flash effects \cite{Jeong:2006, Chen:2010, Chalony:2011, Kwong:2014, Kwong:2015}, the only supplementary ingredient being the scattering event, which does modify the decay rate. Linear-optics superradiance is thus mainly a dispersion effect.

Besides providing a nice physical description, this picture also allows us to derive a simple equation to compute the early decay of scattered light. This `linear-dispersion' (LD) theory, first introduced in \cite{Kuraptsev:2017} and recently used to simulate the excitation dynamics at the switch-on \cite{Guerin:2019,EspiritoSanto:2020}, is very efficient from the computing point of view. We can then apply it to a direct quantitative comparison with experimental data and to the problem of moving atoms \cite{Weiss:2019,Kuraptsev:2020}.

The paper is organized as follows. In the next section we describe the LD theory of superradiance in a very simple way and benchmark it against the more commonly used coupled-dipole (CD) model. In Sec. \ref{sec.exp} we compare its results with experimental data from \cite{Araujo:2016}. In its limit of validity, which is that multiple scattering should be negligible, the agreement is very good. Finally in Sec. \ref{sec.motion} we address the case of thermal motion and show that, in principle, superradiance can be observed with room-temperature vapors.

\section{Linear-dispersion theory of superradiance}\label{sec.theory}

The main physical ingredient of the LD theory is to consider the different frequency components of the field that drives the atoms. These frequency components are due to the switch-on and -off of the field. The LD theory is thus relevant to study nonstationary effects in light-atom interactions, such as the transient behavior at the switch-on \cite{Guerin:2019,EspiritoSanto:2020} and the superradiant decay at the switch-off.

\subsection{Simple version of the linear-dispersion theory}

For pedagogical purposes, we only present here the result of the LD theory without derivation and in its simplest version, which corresponds to the case of motionless atoms and a scalar model for light. We also do not bother with numerical prefactors. A more sophisticated version, accounting for the polarization, Zeeman states and a velocity distribution, is given in the Appendix \ref{sec.A}, along with the main ideas and approximations for its derivation.

Let us start from the result and explain its meaning. The intensity $I_{\bm{k'}}(t)$ detected in the direction $\bm{k'}$ as a function of time $t$ is given by
\begin{multline}
\label{eq.Sokolov}
I_{\bm{k'}}(t) \propto \int d^3\bm{r} \rho(\bm{r}) \, \left| \int_{-\infty}^\infty d\omega E_0(\omega) e^{-i\omega t} \right. \\ \left. \times \exp\left[i\frac{b_0(\bm{r},\bm{k'})}{2}\tilde{\alpha}(\omega)\right] \, \tilde{\alpha}(\omega) \, \exp\left[i\frac{b_0(\bm{r},\bm{k})}{2}\tilde{\alpha}(\omega)\right] \right|^2.
\end{multline}
In this equation, $\rho(\bm{r})$ is the atomic density distribution, $E_0(\omega)$ is the Fourier transform of the incident field,
\begin{equation}\label{eq.alphatild}
\tilde{\alpha}(\omega) = \frac{-1}{i + 2(\omega-\omega_\at)/\Gamma_0} %= \frac{i - 2(\omega-\omega_\at)}{1+4(\omega-\omega_\at)^2/\Gamma^2}
\end{equation}
is the dimensionless atomic polarizability with $\Gamma_0$ the natural linewidth, and the $b_0(\bm{r},\bm{k})$ terms denote the resonant optical thickness through a part of the cloud, from the position $\bm{r}$ into the direction $\bm{k'}$, and from the incident direction $\bm{k}$ to the position $\bm{r}$. In the case of a Gaussian cloud of rms size $R$, and taking the incident wave vector along the $z$ axis and putting the origin of coordinates at the center of the cloud, one can easily compute:
\begin{equation}\label{eq.partial_b0}
\begin{split}
b_0(\bm{r},\bm{k'}) & = \frac{b_0}{2} \exp\left[\frac{-r^2 + (\bm{r}\cdot \bm{k'})^2}{2R^2}\right] \, \left[1- \mathrm{erf}\left(\frac{\bm{r}\cdot \bm{k'}}{\sqrt{2}R}\right) \right] \\
b_0(\bm{r},\bm{k}) & = \frac{b_0}{2} \exp\left[-\frac{x^2+y^2}{2R^2}\right] \, \left[1+ \mathrm{erf}\left(\frac{z}{\sqrt{2}R}\right) \right] \, ,
\end{split}
\end{equation}
where $b_0 = \sqrt{2\pi} \rho_0 \sigma_0 R$ is the resonant optical thickness through the center of the cloud, with $\rho_0$ the peak atomic density and $\sigma_0$ the resonant scattering cross section.

The meaning of Eq.\,(\ref{eq.Sokolov}) is clear: each Fourier component of the initial field propagates through the cloud until the scattering position $\bm{r}$, propagation during which it undergoes attenuation and dephasing following Beer's law. Then it is scattered at position $\bm{r}$ with some probability and associated dephasing given by the atomic polarizability (\ref{eq.alphatild}). Finally it propagates again through the atomic cloud until it escapes the sample. The whole process acts as a linear transfer function, which applies to the frequency components of the incident field. The temporal dependence is recovered by a Fourier transform and the intensity is computed by taking the squared modulus. Then all possible scattering positions are summed up.

This equation is valid for single-scattering only, since there is only one scattering term.
Note also that the average over the scattering positions is done on the intensity: the random phase associated with incoherent scattering and the associated speckle pattern are averaged out. Indeed, what is computed is formally a quantum-mechanical average, i.e., an average over the disorder configurations (see Appendix \ref{sec.A}). Still, since this model describes superradiance, as we show below, it means that superradiance is not related to the interference between light scattered by different atoms. It is actually related to the interference between the different Fourier components of the incident field scattered by the atoms and attenuated or dephased by the surrounding effective medium. It is thus mainly a \emph{dispersion} effect. Of course, the complex refractive index of the effective medium can also be considered as an interference effect between light coherently scattered by all atoms.

This calculation is very similar to what is done to explain the transient effects observed in the coherently transmitted beam, the so-called optical precursors and flash effects \cite{Jeong:2006, Chen:2010, Chalony:2011, Kwong:2014, Kwong:2015}, except for the extra scattering term. In this case, though, this approach is more natural because there is no random phase associated with any scattering. The extra scattering term introduces a quantitative difference between superradiance off-axis, as observed in \cite{Araujo:2016}, and superradiance of the forward scattering lobe, as studied in \cite{Scully:2006,Svidzinsky:2008,Courteille:2010,Roof:2016}.

In this approach, one can understand the occurrence of a superradiant decay rate ($\Gamma_\supp > \Gamma_0$) for large $b_0$ by the spectral broadening of the transfer function induced by the larger value of $b_0$: if the transfer function becomes broader in Fourier space, the temporal response becomes faster. This is also the intuitive picture given for the flash effect, which can also have a decay rate faster than $\Gamma_0$ \cite{Kwong:2015}.

\subsection{Benchmark against the coupled-dipole model}

\begin{figure}
\centering\includegraphics{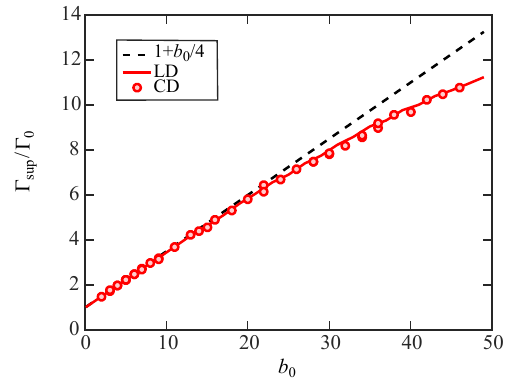}
\caption{Comparison between the coupled-dipole (CD) model and the linear-dispersion (LD) theory for the superradiant decay rate $\Gamma_\supp$. Also shown is the analytical result for the large-detuning limit. The detuning is $\Delta=-10\Gamma_0$, the observation direction is $\theta=45^\circ$. The fitting range is $0<t<0.02\Gamma_0^{-1}$.}
\label{fig.Comparison_CD_Sokolov}
\end{figure}

In Fig.\,\ref{fig.Comparison_CD_Sokolov} we compare the results of the decay rate fitted at very early time on temporal traces computed from the CD model and the linear-dispersion (LD) model (Eq.\,\ref{eq.Sokolov}), at large detuning ($\Delta=-10\Gamma_0$). The agreement is excellent. Note that several densities are used in the CD simulations, from $\rho_0\lambda^3=1.5$ to $25$ (with an exclusion volume $k_0 r_{ij}>0.5$), while the density is not a parameter in the LD model. This shows that, in this range of parameters, the density does not plays any role.

We also show an analytical result that can be computed  from Eq.\,(\ref{eq.Sokolov}) using the residue theorem. For a squared pulse of duration $T \gg \Gamma_0^{-1}$, in the limit $t \rightarrow 0$, i.e., right after the switch-off, and under the condition $b_0 \Gamma_0/\Delta \ll 1$, we obtain (see \cite{Kuraptsev:2017})
\begin{equation}\label{eq.Gamma_Sokolov}
\Gamma_\supp = \left(1 + \frac{b_0}{4} \right) \Gamma_0 \, .
\end{equation}
Note that for isotropic samples, this does not depend on the observing direction, and that the model only contains superradiance off-axis, i.e., with a true scattering event; it does not include the forward lobe of the timed-Dicke (TD) state \cite{Scully:2006,Svidzinsky:2008,Courteille:2010, Bromley:2016, Roof:2016}. As a consequence the decay rate is different, even in the forward direction, than for the TD state. The extra $\tilde{\alpha}(\omega)$ term (scattering) is responsible for a factor 2 in the superradiant enhancement factor. It emphasizes the different nature of the forward lobe, which is, in a photon picture, diffracted or refracted light by the effective medium, without any true scattering. The different superradiant decay rates between on-axis and off-axis scattering was already numerically observed in \cite{Araujo:2016}.

A deviation from this analytical limit can be seen for the largest $b_0$'s. This is the first sign of the suppression of superradiance as soon as $b_0 \Gamma_0/ \Delta$ is not small. A systematic numerical study shows that the superradiant decay rate reaches it maximum at $b_0 \simeq 8\Delta/\Gamma_0$ and decays beyond this value. Then, for very large $b_0$, it slowly tends toward $\Gamma_0$, but the LD model is not valid in that regime because the actual optical thickness $b(\Delta) = b_0/(1+4\Delta^2/\Gamma_0^2)$ is not small and thus multiple scattering is not negligible any more.
A possible interpretation for the decrease of $\Gamma_\supp$ is that the initial detuning must be much larger than the width of the transfer function. Then the incident spectrum $E_0(\omega)$ is almost constant at the scale of the transfer function and thus does not narrow the transferred spectrum, yielding the fastest dynamics. This leads to the condition $\Delta \gg \Gamma_\supp/2$, which corresponds well to $b_0 \ll 8\Delta/\Gamma_0$. Note that another, but consistent, interpretation of the reduction of superradiance close to resonance has already been discussed in the framework of the CD model \cite{Guerin:2017b}.

%\WG{Maybe one should make a dedicated numerical study on the range of validity of the LD approach, via systematic comparison with the CD for different values of $b_0$, $\Delta$. And also, using the LD only, to determine more precisely when the decay rate starts to drop. Such a study would greatly help the interpretation of the results in the following parts.}

%Actually, while doing this comparison, we observed that the good agreement between the two models, as well as the deviation from Eq.\,(\ref{eq.Gamma_Sokolov}), was quite sensitive to the fitting range used for determining the decay rate. The reduction of the superradiant decay rate at large $b_0$ is less visible if the decay rate is measured closer to the switch-off time $t=0$: The superradiant decay rate measured right at the switch-off is more robust, but it is more and more limited in time as the detuning goes closer to resonance, and thus harder to see. This is consistent with a decrease of the superradiant population when the driving field is near resonance, as discussed in \cite{Guerin:2017b}.

Of course, the LD theory is extremely efficient from a computing point of view. Moreover there is no limitation for the atom number or $b_0$ (contrary to the CD model), and we can also include the Zeeman structure. It is thus possible to make a direct comparison with experimental data without any free parameter. Another possible extension is to include the effect of atomic motion, which can be done by a simple Doppler broadening of the atomic polarizability and a Doppler shift at the scattering (see Appendix \ref{sec.A}). Whereas the CD simulations with moving atoms are extremely time demanding \cite{Weiss:2019,Kuraptsev:2020}, the LD theory allows fast computing. We address those two problems in the following.

\section{Comparison with experimental data}\label{sec.exp}

\begin{figure}[b]
\centering\includegraphics{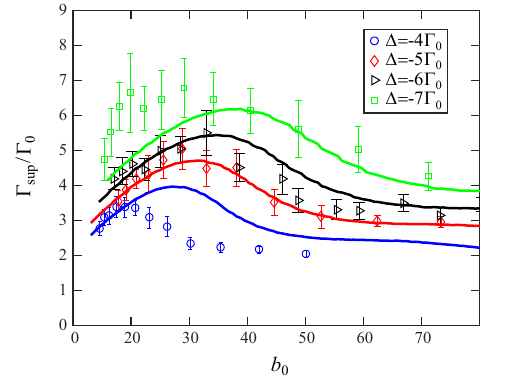}
\caption{\textit{Ab initio} comparison between the experimental data (symbols) from \cite{Araujo:2016} and the linear-dispersion theory (solid lines) for the superradiant decay rate $\Gamma_\supp$ as a function of the resonant optical thickness $b_0$ for different detunings. Here the fitting range starts as $t>0.1 \Gamma_0^{-1}$ (to wait for the laser switch-off) until the detected light intensity decreases to $20\%$ from its steady-state value (before the switch-off).}
\label{fig.Comparison_exp}
\end{figure}

To perform such a comparison we can use the experimental results of Ref. \cite{Araujo:2016} on off-axis superradiance \footnote{Note that in  \cite{Araujo:2016}, the decay rates were slightly underestimated by the fact that a portion of the detected scattered light was coming from a hot-vapor background. We have refined the analysis to remove this systematic effect in the decay rates reported in Fig.\,\ref{fig.Comparison_exp}.}.
For the linear-dispersion modeling, we have simulated as closely as possible the experimental switch-off profile of the laser, which has been measured independently \footnote{Since the pulse profile is different, a direct comparison with the results of the previous section is not possible.}. About the multilevel aspect of the rubidium atoms used in the experiment, it actually does not change the modeling. Indeed, we suppose that all Zeeman states are equipopulated and, in addition, the total light intensity was measured without any polarization selection \footnote{There might be some spurious polarization selection introduced by the optical elements between the atoms and the detector but we suppose here that this is negligible.}. Under these conditions, one can show that the complete multilevel equation (\ref{25}) is equivalent to Eq.\,(\ref{eq.Sokolov}), where $b_0$ is the resonant optical thickness measured in the experiment, which includes a degeneracy factor in the scattering cross-section.

The result of the LD theory is reported in Fig.\,\ref{fig.Comparison_exp} along with the data points. The decay rates are determined by exponential fits using the same fitting window for both (see caption). Except for the lowest part of the $\Delta=-7\Gamma_0$ data set, where there is a small discrepancy that is not understood, the agreement is very good for large detunings, without any free parameter. Yet a discrepancy appears at large $b_0$ for the lowest detuning $\Delta=-4\Gamma$. Other data sets at even lower detuning (not shown for clarity) are also not in agreement with the LD theory. We attribute this discrepancy to multiple scattering, which is neglected in the LD theory: as soon as the condition $b(\Delta)\ll1$ is not fulfilled any more, deviations from the LD theory start to appear. For $\Delta=-4\Gamma_0$, the discrepancy starts at $b_0>20$, corresponding to $b(\Delta)>0.3$. We note, however, that for data at larger $\Delta$, the agreement seems to go beyond this limit.

Interestingly, one can also notice that some of the measured and computed decay rates are larger than the prediction of Eq.\,(\ref{eq.Gamma_Sokolov}). Using different switch-off profiles in the LD theory, we have checked that an instantaneous switch-off, which leads to Eq.\,(\ref{eq.Gamma_Sokolov}), does not produce the fastest decay rate, which is somewhat counterintuitive. One possible explanation is that resonant photons produce longer-lived excitations \cite{Lagendijk:1996, Bourgain:2013}. For an initially detuned field, a large broadening (fast extinction) is thus not favorable for superradiance. This effect will be the subject of further investigations.

Finally, the large-$b_0$ limit is not visible in Fig.\,\ref{fig.Comparison_exp} but we have checked that the LD theory predicts a decay rate slowly reaching the single-atom one (e.g., $\Gamma_\supp \simeq 1.14 \Gamma_0$ for $\Delta=-4 \Gamma_0$ and $b_0=200$), although the experiment yields lower values due to multiple scattering \cite{Araujo:2016}.

\section{Superradiance with thermal motion}\label{sec.motion}

One can easily include the effect of atomic motion in the LD theory by convoluting the atomic polarizability with the detuning distribution corresponding to the Doppler broadening (Eq.\,\ref{25}) The Doppler broadening also acts on the propagation part (Eq.\,\ref{24}) and one should also take into account the Doppler shift between the incident light and the scattered light. For the computing time, it is tremendously more efficient than solving the CD equations with moving atoms, as done in \cite{Weiss:2019,Kuraptsev:2020} for subradiance.

\subsection{Benchmark against the coupled-dipole model}

\begin{figure}[t]
\centering\includegraphics{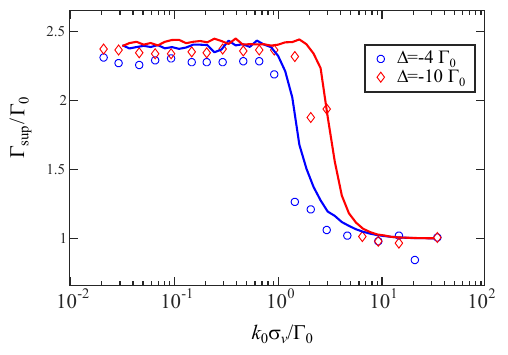}
\caption{Comparison between the coupled-dipole model (symbols) and the linear-dispersion theory (solid lines) for the superradiant decay rate $\Gamma_\supp$ as a function of the normalized Doppler width $k_0 \sigma_v / \Gamma_0$ ($\sigma_v$ is the rms width of the velocity distribution), for two detunings, $\Delta=-4$ (blue) and $\Delta = -10$ (red). The parameters are $N=1500$, $\rho_0\lambda^3=2$, $b_0=5.78$, $\theta=\pi/2$, averaged over $\varphi$ and over $\sim 200$ realizations, and the fitting range is $0<t<0.15 \Gamma_0^{-1}$.}
\label{fig.Comparison_CD_LD_temp}
\end{figure}

We use the same CD simulation method as in \cite{Weiss:2019}, with moving atoms. The comparison between the two models is shown in Fig.\,\ref{fig.Comparison_CD_LD_temp} for a fixed $b_0$ and two different detunings. The agreement is good, with only a slight discrepancy in the $T=0$ limit. This discrepancy was absent in Fig.\,\ref{fig.Comparison_CD_Sokolov} computed with motionless atoms. In that case we used an exclusion volume to suppress the influence of strong superradiant pairs, while we removed the exclusion volume for the case of moving atoms. These superradiant pairs (with at most $\Gamma=2$) contribute to a slight decrease of the superradiant decay rate in the CD model.

The two models agree well in predicting a certain robustness of superradiance: the superradiant decay rate is unaffected until a Doppler broadening on the order of $\Gamma_0$. For $^{87}$Rb, it corresponds to $T\sim 235$\,mK, well above standard temperatures of cold-atom experiments. In Appendix \ref{sec.B} we show experimental data that confirms this robustness until $T\sim 11$\,mK.

The two models also agree in their prediction of suppressed superradiance at higher temperature, with a critical temperature that clearly depends on the detuning: a larger detuning allows one to observe superradiance at larger temperature. We interpret this behavior by the reduction and suppression of superradiance when the driving field is close to resonance, as discussed in the previous sections (and in \cite{Araujo:2016, Guerin:2017b}). Here, the same effect occurs with the Doppler-broadened resonance.

This raises the question of the possibility of observing superradiance at higher temperature, even at room temperature, by using very large detunings.
When solving the CD equations, larger detunings need finer time sampling, which increases the computation time. Detunings larger than the Doppler broadening are thus very difficult to explore with the CD model at large temperature.

\subsection{Room-temperature superradiance}

\begin{figure}[t]
\centering\includegraphics{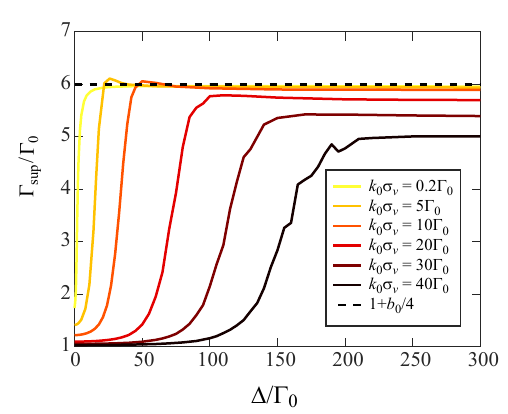}
\caption{Superradiant decay rate computed from the LD theory as a function of the detuning for different values of the Doppler broadening, with $b_0=20$. The decay rate is determined in the range $0<t<0.02\Gamma_0^{-1}$. At large detuning $\Delta \gg k_0 \sigma_v$, superradiance is recovered.}
\label{fig.High_T}
\end{figure}

In Fig.\,\ref{fig.High_T} we show the superradiant decay rate computed from the LD theory as a function of the detuning for several values of the  Doppler broadening, up to $k_0\sigma_v = 40 \Gamma_0$, corresponding to $T\approx 400$\,K for Rb. Correspondingly, the detuning goes up to $\Delta = 300 \Gamma_0$. The resonant optical thickness is fixed, $b_0=20$ (defined for motionless atoms). As expected, superradiance is suppressed near resonance. At very large detuning, however, superradiance is recovered.

With some algebra, it is actually possible to analytically show, starting from Eq.\,(\ref{25}), which includes the atomic velocity distribution, that one should recover the motionless superradiant decay rate in the limit $\Delta \gg k_0\sigma_v$. Numerically we obtain a slight discrepancy, which is due to the fitting range (see caption): the maximum decay rate is only visible for a very short time after $t=0$, beyond our numerical resolution. Note also that the decay rate is overestimated for the lowest temperatures on resonance because multiple scattering should be significant there. % On the other hand, the chosen fitting range is short enough such that on resonance, and for the lowest temperature, superradiance is not completely suppressed ($\Gamma_\supp>1$).
%\WG{Could we try to quantify how the superradiant weight evolves...?} \WG{Also show $\Gamma_\supp$ vs $k_0\sigma_v$ for $\Delta=0$ and $\Delta=\infty$ ? Look at scaling law, compare with CBS decoherence???}

From these results, it seems possible to observe superradiance in the linear-optics regime using room-temperature vapor at moderate optical thickness. Since the driving field must be largely detuned, it interacts only very weakly with the atomic vapor: therefore the experimental difficulty is to collect enough scattered light (compared to spurious light).

\section{Conclusion}

We have presented a linear-dispersion model for off-axis superradiance in the linear-optics regime for disordered dilute samples. This approach is very useful from a computational point of view, as it provides a fast method of calculating the superradiant dynamics. It also gives a nice description of the physics from the point of view of light. Superradiance appears as a dispersion effect, similar to optical precursors, without involving interference between light scattered by different atoms.

If one uses Eq.\,(\ref{eq.Sokolov}) to compute the decay at all time, one can check that after the fast superradiant part, the following of the decay tends to a single exponential of rate $\Gamma_0$: the LD theory does not describe subradiance. This means that linear-optics subradiance cannot be understood as a dispersion effect, contrary to linear-optics superradiance. An optical description of subradiance in disordered samples is therefore still an open problem. It should rely on mechanisms not included in the LD approach, such as multiple scattering \cite{Labeyrie:2003, Weiss:2018}, recurrent scattering or refractive-index-gradient trapping \cite{Schilder:2016, Cottier:2018}. Which of these mechanisms is the dominant one probably depends on the parameters of the experiment, such as the detuning of the excitation and the density of the sample \cite{Fofanov:2021,Cipris:2021b}. Note that in multiple-scattering approaches the effective medium between scattering events also plays an important role \cite{Labeyrie:2003, Weiss:2018, Labeyrie:2002}.

Another open problem is the extension of this model to the case of dense samples. At high density other phenomena occur, such as collective shifts and recurrent scattering \cite{Javanainen:2016}, which are not included in the present model. However it might be possible to keep the essential ingredients of the linear-dispersion theory and include the high-density effects through a renormalization of the atomic susceptibility, following the method recently presented in \cite{Andreoli:2020}.

\section*{Acknowledgements}

W.\,G. thanks Romain Pierrat for initial help.
Part of this work was performed in the framework of the European Training Network ColOpt, which is funded by the European Union (EU) Horizon 2020 program under the Marie Sklodowska-Curie action, Grant Agreement No. 721465, and of the project ANDLICA, ERC Advanced Grant No.\,832219. We also acknowledge funding from the French National Research Agency (Projects PACE-IN ANR19-QUAN-003 and QuaCor ANR19-CE47-0014). P.\,W. received support from the Deutsche Forschungsgemeinschaft (Grant No. WE 6356/1-1). I.\,M.\,S. thanks the Ministry of Science and Higher Education of the Russian Federation for financial support (the State Program for Fundamental Research, theme code FSEG-2020-0024).

\appendix

\section{Main steps for the derivation of the linear-dispersion theory} \label{sec.A}

%\WG{This is the shortest version (1.5 pages) of Igor's Appendix. We also have a 3-page version and a 4.5-page version. To be discussed what is best... Igor, I cleaned up your text, please check if I didn't introduce any mistake.}

The intensity $I_\nu(\mathbf{\Omega},t)$ of the light polarization component $\nu$ that the atomic ensemble scatters
in a unit solid angle around an arbitrary direction given by the vector $\mathbf{R}$  ($\mathbf{\Omega}=(\theta,\varphi))$ is determined by the electric field second-order correlation function $D_{\nu _{_{1}}\nu_{_{2}}}^{(E)}(\mathbf{r}_{1},t_{1};\mathbf{r}_{2},t_{2})$ via \cite{Glauber:1965}
\begin{equation}
I_\nu(\mathbf{\Omega},t)=\frac{c}{2\pi }
\int\limits_{S_{u}}\!d^{2}r D_{\nu\nu}^{(E)}(\mathbf{r},t;\mathbf{r},t). \label{1.1}
\end{equation}
Here the integral is calculated over a spherical surface $S_{u}$ corresponding to a unit spherical angle and located far from the considered atomic ensemble. The center of the sphere is assumed to be in this ensemble and $c$ is the speed of light in vacuum.

In order to theoretically describe the effect of superradiance, we have to be able to calculate the correlation function $D_{\nu\nu}^{(E)}(\mathbf{r},t;\mathbf{r},t)$. In the general case this function is
expressed in terms of negative-frequency $E_{\nu _{_{1}}}^{(-)}(\mathbf{r},t)$ and
positive-frequency $E_{\nu _{_{1}}}^{(+)}(\mathbf{r},t)$ components of the
Heisenberg electric field operators:
\begin{equation} D_{\nu
_{_{1}}\nu
_{_{2}}}^{(E)}(\mathbf{r}_{1},t_{1};\mathbf{r}_{2},t_{2})=\langle
E_{\nu _{_{2}}}^{(-)}(\mathbf{r}_{2},t_{2})E_{\nu
_{_{1}}}^{(+)}(\mathbf{r}_{1},t_{1})\rangle . \label{2}
\end{equation}%
The brackets in this expression correspond to
quantum-mechanical statistical averaging over the density
operator of the entire system under investigation.

The correlation function (\ref{2}) can be calculated by the diagram technique for nonequilibrium systems (see for example \cite{Lifshitz:1981, Konstantinov:1961, Keldysh:1965, Datsyuk:2006, Kupriyanov:2017}). This technique is based on a perturbation-theory expansion. The field correlation function is expanded into series over interaction between atoms and light. Each item in this expansion is represented by a diagram. Part of the diagrams can be summed up.

In this paper we consider only the case of side scattering, when the mean values of the field operators are equal to zero, $\langle E_{\nu}^{(\pm)}(\mathbf{r,}t)\rangle  =0$, in the region of the photodetector. In addition, when calculating the correlation function (\ref{2}), we assume the following typical experimental conditions: The initial states of the atomic ensemble and light are uncorrelated; the atomic ensemble is dilute, which means that the average interatomic distance is much larger than the wavelength of the resonant radiation.

The exciting light is assumed to be a long pulse and the time profile of its positive frequency component can be determined by the following superposition of monochromatic waves
\begin{equation}
{\cal E}_\mu^{(+)}(\mathbf{r},t)=u_\mu\int\limits_0^\infty\frac{d\omega}{2\pi}{\cal E}(\omega)\exp(i\mathbf{kr}-i\omega t),
\label{10.1}
\end{equation}
where the unit vector $\mathbf{u}_\mu$ determines the polarization of the incident light. This light is weak and all nonlinear optical phenomena can be neglected.

Under such assumptions we can get an expansion of the correlation function (\ref{2}) as a series over the number of events
of incoherent scattering of a photon in the medium. By incoherent, in contrast to coherent forward scattering, we mean an act of scattering of a photon by an atom in which the direction of the wave vector changes. Coherent forward scattering can be taken into account at all orders by introducing the exact advanced and retarded Green's functions of the electromagnetic field in the considered medium. They can be found analytically as solutions of the corresponding Dyson equation (see for example \cite{Datsyuk:2006, Kupriyanov:2017}). In a similar way we can sum up all diagrams responsible for the interaction of excited atoms with the vacuum reservoir and introduce `dressed' atomic Green's functions. They also can be found analytically \cite{Datsyuk:2006, Kupriyanov:2017}.

Every event of incoherent scattering leads to a delay of the secondary photon caused by the so-called dwell or Wigner time \cite{Lagendijk:1996, Bourgain:2013}. For this reason, to describe the process of superradiance, which takes place just after the exciting pulse is switched off, we can consider only the contribution of a single incoherent scattering event.

Omitting all calculations, which can be found in \cite{Datsyuk:2006, Kupriyanov:2017}, we reproduce here the final result for the single scattering contribution $I^s_\nu(\mathbf{\Omega},t )$ in the light intensity $I_\nu(\mathbf{\Omega},t)$. This contribution  is
\begin{gather}
I^s_\nu(\mathbf{\Omega},t )=\int\!\! d^3 r\int \!\!\frac{d^{3}p}{(2\pi \hbar
)^{3}} \frac{c}{4\pi\hbar^2}\sum\limits_{m}\rho _{mm}(\mathbf{p,r}) \notag\\
\times \left\vert\sum\limits_{\mu',\nu',m'}\int\limits_{0}^{\infty }
\dfrac{k^{2}d\omega }{2\pi }{\cal E}(\omega) \right. \exp(-i\omega t)\times \label{25} \\
\left. \mathbf{u}^{\prime}_{\nu}X_{\nu\nu'}(\mathbf{R} ,\mathbf{r},\omega
^{\prime })\alpha _{\nu'\mu'}^{(m^{\prime}m)}(\omega -\mathbf{kv})
X_{\mu'\mu}(\mathbf{r}, \mathbf{r}_0,\omega )\mathbf{u}_{\mu}  \right\vert ^{2}. \notag
\end{gather}
This is valid even for anisotropic media when the populations of the different Zeeman sublevels are different.
Here the unit vectors $\mathbf{u}^{\prime}$  correspond to the two possible orthogonal
polarizations of scattered light; the function $X_{\mu'\mu}(\mathbf{r}, \mathbf{r}_0,\omega )$ describes the propagation of light
from the source $\mathbf{r}_0$ to the point $\mathbf{r}$, where a single incoherent
scattering event takes place. The function $X_{\nu\nu'}(\mathbf{R} ,\mathbf{r},\omega
^{\prime })$ describes the propagation of a secondary photon with frequency $\omega'$ toward the photodetector. The frequency $\omega'=\omega+(\mathbf{k}'-\mathbf{k})\cdot \mathbf{v}$ differs from $\omega$ due to a Doppler shift.

In the typical case of isotropic media, when all Zeeman sublevels of the ground state are uniformly populated and the single atom density matrix
$\rho _{mm}(\mathbf{p},\mathbf{r})$ does not depend on the magnetic quantum number $m$, the function $X_{\mu _{_{1}}\mu_{_{2}}}(\mathbf{r}_{1},\mathbf{r}_{2};\omega
)$ can be calculated as follows:
\begin{equation}
X_{\mu _{_{1}}\mu_{_{2}}}(\mathbf{r}_{1},\mathbf{r}_{2};\omega
)=\exp(-b(\mathbf{r}_{1},\mathbf{r}_{2},\omega)/2)
\delta_{\mu _{_{1}}\mu_{_{2}}} \, , \label{23}
\end{equation}
where the `complex optical thickness' (accounting for attenuation and dephasing) of the inhomogeneous
cloud between points $\mathbf{r}_{1}$ and $\mathbf{r}_{2}$ for the considered case is
\begin{gather}
b(\mathbf{r}_{1},\mathbf{r}_{2},\omega)=\frac{4\pi k' \|d_{j_0j}\|^2}{3\hbar}\int\limits_
{\mathbf{r}_{2}}^{\mathbf{r}_{1}} n(\mathbf{r})ds\notag \\
\times\int \!\frac{d^{3}p}{{(2\pi \hbar
)}^{3}}\frac{f(\mathbf{p})}{-i(\omega -\omega
_{jj_0}-\mathbf{k'v})+\Gamma_0/2}.  \label{24}
\end{gather}
Here $n(\mathbf{r})$ and $f(\mathbf{p})$ are the spatial and momentum distributions of atoms in the considered ensemble, $\|d_{j_0j}\|$ is the reduced matrix element of the dipole operator for the transition between the ground and excited states of total angular momentum $j_0$ and $j$ respectively, and $\Gamma_0$ is the natural linewidth. The wave vector $\mathbf{k'}$ is directed along $\mathbf{r}_{1}-\mathbf{r}_{2}$.

\begin{figure*}[t]
\centering\includegraphics{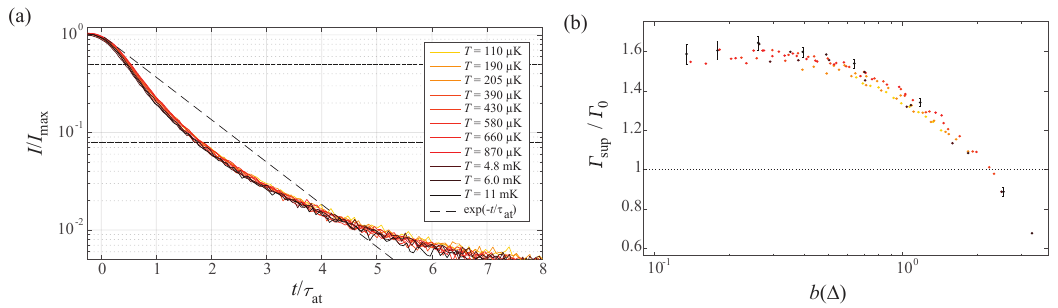}
\caption{Experimental superradiance data for $\Delta = -4 \Gamma_0$. (a) Decay curves for light scattered off axis with $b_0 = 35$. The amplitude is normalized to $1$ for the steady-state level, right before the switch-off at $t=0$. The dashed line is the single-atom decay and the temperature is encoded in the color. The two horizontal dashed lines indicate the range used for fitting the decay rates. (b) Fitted decay rates as a function of $b(\Delta)$ for different temperatures (same color code). For clarity, statistical error bars are shown for one data set only.}
\label{fig.data_temp}
\end{figure*}

The matrix $\alpha _{\nu\mu }^{(m^{\prime }m)}(\omega
)$ is a scattering amplitude of the probe photon on an atom:
\begin{equation}
\alpha _{\nu\mu }^{(m^{\prime }m)}(\omega
)\;=\;-\sum_{n}\,\frac{(d_{\nu})_{m^{\prime
}m}\,(d_{\mu })_{nm}}{\hbar (\omega -\omega _{nm})\,+\,i\hbar
\Gamma_0/2}\,.  \label{26}
\end{equation}

Note that expression (\ref{25}) describes only the contribution of single incoherent scattering. It can be used for the description of superradiance when the average optical thickness of the cloud is small. Another restriction is that Eq.\,(\ref{25}) can be used for all directions except in the zones of backward and forward scattering. For forward scattering the main contribution comes from the coherent component of the scattered light, and for backward direction, one of the polarization components is absent for single scattering and scattering of higher order should be taken into account. Equation (\ref{25}) is also not valid for a cloud with a large aspect ratio. In such a case diffraction and refraction effects play essential roles \cite{Roof:2015, Sutherland:2016, Saint-Jalm:2018} and the propagation function $X$ cannot be described by Eq.\,(\ref{23}).

\section{Experimental data on superradiance as a function of the temperature} \label{sec.B}

%\WG{To be discussed if we want to put those data or not. They are of limited interest because they show no dependence in the explored range (so they show robustness) and because the switch-off was not very fast, so superradiance is limited. But we have them and it's the only opportunity to publish them. It fits the topic of section IV. It's clearly not necessary but as appendix, it doesn't harm. What do you think?}

In this appendix we show some experimental data on the superradiant decay as a function of the temperature of the sample. The experimental setup and procedure are the same as in Ref.\,\cite{Weiss:2019} devoted to the influence of atomic motion on subradiance. Here we analyze the temporal dynamics of the decay of scattered light at early time.

Several decay curves are shown in Fig.\,\ref{fig.data_temp}(a) for fixed optical thickness and detuning and for temperatures between 110\,$\mu$K and 11\,mK, corresponding to normalized Doppler broadening $k_0 \sigma_v/\Gamma_0$ between 0.022 and 0.22. All curves exhibit a superradiant behavior, with an early decay significantly faster than the single-atom decay. There is no visible difference between the data acquired at different temperatures, which demonstrate the robustness of superradiance in this temperature range. This is confirmed in Fig.\,\ref{fig.data_temp}(b), where we show the fitted superradiant decay rate as a function of the optical thickness for the same temperatures. At the precision of the experiment and in this limited range of temperature, we do not observer any influence of the temperature. The behavior is the one observed in Fig.\,\ref{fig.Comparison_CD_Sokolov} for motionless atoms, i.e. an increase with the optical thickness as long as $b(\Delta) \ll 1$, and then a decrease. Note also that superradiance is more limited than in the data of Fig.\,\ref{fig.Comparison_exp} because the switch-off of the laser was significantly slower ($\sim 15$\,ns instead of $\sim 3$\,ns).

%%\bibliographystyle{apsrev4-1}
%\bibliography{D:/RECHERCHE/MesPublis/MaBiblio/AllMyBiblio} % office PC
%%\bibliography{F:/Backup_PC_bureau/RECHERCHE/MesPublis/MaBiblio/AllMyBiblio}  % external hard drive

%merlin.mbs apsrev4-1.bst 2010-07-25 4.21a (PWD, AO, DPC) hacked
%Control: key (0)
%Control: author (0) dotless jnrlst
%Control: editor formatted (1) identically to author
%Control: production of article title (0) allowed
%Control: page (1) range
%Control: year (0) verbatim
%Control: production of eprint (0) enabled
%

\end{document}